\documentclass[letterpaper,twocolumn,prl]{revtex4-1}
\usepackage[latin9]{inputenc}
\usepackage{verbatim}
\usepackage{mathtools}
\usepackage{amsmath}
\usepackage{amssymb}
\usepackage{graphicx}
\usepackage{wasysym}

\makeatletter

\pdfpageheight\paperheight
\pdfpagewidth\paperwidth

\usepackage{lipsum}
\usepackage{dcolumn}
\usepackage{bm}
\usepackage{bm}
\usepackage{braket}
\usepackage{color}
\usepackage{float}
\usepackage{subfigure}
\usepackage{color}
\usepackage{soul}
\usepackage{sidecap}
\usepackage{float} 
\newcommand{\re}{\textcolor{red}}

\definecolor{violet}{rgb}{0.58, 0.0, 0.83}

\newcommand{\mkaddcomment}[1]{{\bf {\color{violet}{[MK: #1]}}}} 

\newcommand{\rcadd}[1]{{\color{blue}\bf{#1}}}

\definecolor{red}{rgb}{0.68, 0.0, 0.13}

\makeatother

\begin{document}
\title{Cavity induced many-body localization}
\author{Rong-Chun Ge$^{1,2}$, Saeed Rahmanian Koshkaki$^{1}$, Michael H. Kolodrubetz$^{1}$}
\affiliation{$^1$Department of Physics, The University of Texas at Dallas, Richardson,
Texas 75080, USA}
\affiliation{$^2$College of Physics, Sichuan University, Chengdu China, 610064}
\begin{abstract}
In this manuscript, we explore the feasibility of achieving many-body localization in the context of cavity quantum electrodynamics at strong coupling. Working with a spinless electronic Hubbard chain sitting
coupled to a single-mode cavity, we show that the global coupling between electrons and photons -- which generally would be expected to delocalize the fermionic excitations -- can instead favor the appearance of localization. This is supported by a novel high-frequency expansion that correctly accounts for electron-photon interaction at strong coupling, as well as numerical calculations in both single particle and many-bod regimes. We find evidence that many-body localization may survive strong quantum fluctuations of the photon number by exploring energy dependence, seeing signatures of localization down to photon numbers as small as $n\sim2$. 
\end{abstract}
\maketitle

{\allowdisplaybreaks
\section{Introduction}

With the dramatic advance in coherent control during the past decade~\cite{COH1,COH2},
an increasing set of controllable quantum systems are available, such
as ultracold atoms, trapped ions, and superconducting qubits. In these
systems, experimentalists can steer the dynamics of the system far
away from thermal equilibrium, which has stimulated a surge of interest
in non-equilibrium dynamics and phases of quantum matter. Examples
include new families of topological phases without equilibrium analogs
\cite{Rudner2013,Roy2016,MBLTOPO2,Else2016,Potter2016,Nathan2015,Roy2017,Titum2016,Po2016,Potter2017,Harper2017,Roy2017a,Po2017,Potter2018,Reiss2018,Kolodrubetz2018,Nathan2019,Timms2021,Long2021,Nathan2021},
novel forms of symmetry breaking such as quantum time crystals \cite{Khemani2016,MBLTOPO3,Else2016a,DFTC3,zhang2017observation,Else2020,KhemaniArxiv2019,Mi2022},
and long-live prethermal states with non-trivial dynamics \cite{RubioAbadal2020,Peng2021,Beatrez2021}.
While in general these systems are fragile and eventually heat to
a featureless infinite-temperature state, there is strong evidence
that this can be circumvented by mitigating slow resonant heating
\cite{Bukov2016a,Weidinger2017}, introducing constraints \cite{Turner2018,Nandkishore2019},
or adding sufficient disorder to realize many-body localization (MBL)~\cite{FMBL1,FMBL2,FMBL3,ResonanceMBL}.

While MBL is a paradigmatic method for avoiding thermalization, its
potential beyond locally interacting models with finite on-site Hilbert
space remains to be understood. Recently, we have provided strong
numerical evidence that MBL can be robust in the context of cavity
quantum electrodynamics (QED), where the local interactions within
a system of particles or spins competes against global interactions
induced by the cavity \cite{MBLCentral1,MBLCentral2,Ng2021}; these results are confirmed elsewhere~\cite{Kubala2021}. These
papers hinged on the well-explored case of MBL in the presence of
time-periodic (Floquet) drive, which becomes a cavity QED model in
the appropriate limit. Extending beyond the Floquet limit involves
a high-frequency expansion, whose (asymptotic) convergence rests on the separation of scales between the weak cavity/atom coupling and the large photon energy scale.

Yet the most interesting Floquet phenomenon involves resonant driving, for which no such separation of scales exists. In this paper, we investigate the possibility of achieving MBL in a strongly coupled model of cavity QED. Intuitively, the presence of a global cavity mode will induce all-to-all couplings, which will be detrimental to achieving MBL. However, we find the presence of the cavity mode instead \emph{favors} localization. We show that this behavior is related to the Floquet phenomenon of coherent destruction of tunneling \cite{Grossmann1991,Lignier2007} and comment on the role of gauge freedom in determining this limit. We provide numerical evidence of a parameter regime where localization survives to a small photon number $n\sim2$ where quantization of the cavity photon is clearly relevant. We find a version of the high-frequency expansion that can be used to understand these strongly and resonantly driven systems, which we expect will be extensible to a wider variety
of cavity QED systems.

\section{Setup and model\label{sec:setup}}

We are interested in the setup illustrated in Figure \ref{fig:setup},
in which the electric field of a single-mode photonic cavity is coupled
to electrons in a one-dimensional system such as a nanowire. The wire is oriented along the polarization direction of the electric field,
such that the electronic polarization of the wire, $P=qd\sum_{j}jn_{j}$
-- where $q$ is the charge of the particles and $d$ is the lattice
spacing -- couples directly to the cavity electric field as $\mathbf{P}\cdot\mathbf{E}$. We assume the wire is much shorter than the waist of the cavity mode, such that the electric field amplitude can be treated as uniform along the length of the wire. If we think of the cavity mode in its semi-classical limit, then the electric field will be $E=E_{0}\cos(\omega t)$ where $\omega$ is the cavity frequency. In the strong coupling limit, this large oscillatory electric field interferes with electronic tunneling, resulting in the phenomenon of coherent destruction of tunneling \cite{Grossmann1991,Lignier2007}. We will be interested to see how this phenomenon holds up in the quantized photon limit, as well as study the role of electronic interactions in localizing the electrons.

\begin{figure}
\centering \includegraphics[clip,width=0.99\columnwidth]{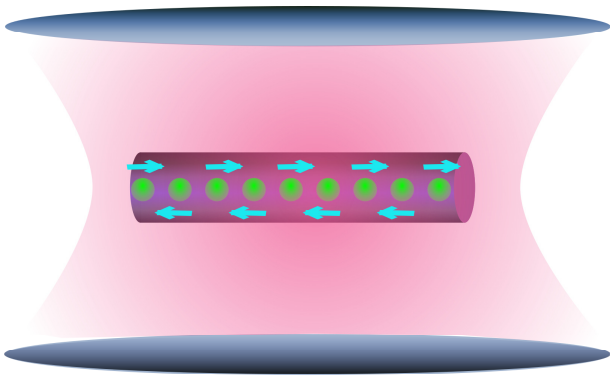}
\caption{Schematic showing a spinless fermion chain inside a photonic cavity.
The electric field polarization of the cavity mode is chosen to be
parallel to the length of the chain, such that photons maximally affect
the particle motion along the wire.}
\label{fig:setup} 
\end{figure}

Modeling this system is surprisingly challenging due to the apparent
gauge ambiguity that has been found in cavity QED systems at strong
coupling \cite{gauge3}. Recently, the gauge ambiguity was found to be induced by truncating the Hilbert space of the matter degrees of freedom~\cite{gauge,gauge1,gauge2,gauge3}.
By appropriate truncations, the physical gauge freedom is restored,
allowing an unambiguous description of the dynamics within a truncated Hilbert space. While many of these ideas came out of the study of cavity-mediated chemical reactions, where the matter consists of interacting molecules, they can readily be adapted to other systems. Here, we use these techniques to model a single band of spinless fermions in our nanowire. We will refer to these spinless fermions as ``electrons'' throughout the majority of this paper, but note that none of our results depend on the particular choice of charged fermion.

Let's begin by considering the most general case, as discussed in
\cite{gauge}, for later generalization. We start with the minimum
coupling Hamiltonian for many fermionic degrees of freedom in a single-mode
cavity (with $\hbar=1$ throughout) in the Coulomb gauge: 
\[
H_{C}^{gen}=\sum_{j}\frac{({\bf p_{j}}-q{\bf A({\bf r}_{j}}))^{2}}{2m}+V({\bf r_{j}})+\omega a^{\dagger}a,
\]
where ${\bf p}_{j}$ and ${\bf r}_{j}$ are the momentum
and position operators of the $j$th particle, $V({\bf r_{j}})$
describes interactions between the particles and their environment
(e.g., with the crystal lattice that will give our band structure),
and $a^{\dagger}$ is the creation operators of the photon mode
with frequency $\omega$. The photon field operator is given by ${\bf A}({\bf r})={\bf f}({\bf r})(ia^{\dagger}-ia)$
with ${\bf f}({\bf r})$ determined by the mode profile. Going into
the dipole gauge using the Power-Zienau-Woolley gauge transform $U_{{\rm PZW}}=\exp\big(-i\sum_{j}q{\bf r}_{j}\cdot{\bf A}({\bf r}_{j})\big)$,
we get~\cite{gauge} 
\begin{align*}
H_{D}^{gen}= & \sum_{j}\left[\frac{p_{j}^{2}}{2m}+V({\bf r_{j}})-\omega q{\bf r}_{j}\cdot{\bf f}({\bf r}_{j})\left(a+a^{\dagger}\right)\right]\\
 & \;\;\;+\omega\left(\sum_{j}q{\bf r}_{j}\cdot{\bf f}({\bf r}_{j})\right)^{2}+\omega a^{\dagger}a,
\end{align*}

The dipole gauge makes approximating this Hamiltonian much easier,
as the electric field has been removed from the kinetic energy term.
We are interested in a simple model of one-dimensional spinless fermions
on a lattice, for which we choose the disordered single-band Hubbard
model. Furthermore, we will assume that the field strength is uniform
across the length of the wire, such that $\mathbf{f}(\mathbf{r}_{j})=\mathbf{f}_{0}$
is constant. For a one-dimensional lattice with lattice spacing $d$
and polarization $f$ parallel to the wire, we see that 
\begin{align}
H_{D} & =\sum_{j}\Bigg[-J\left(c_{j}^{\dagger}c_{j+1}+h.c.\right)+Un_{j}n_{j+1}+V_{j}n_{j}\nonumber \\
 & \;\;\;\;\;\;-\omega\eta jn_{j}\left(a+a^{\dagger}\right)\Bigg]+\omega a^{\dagger}a+\omega\left(\sum_{j}\eta jn_{j}\right)^{2}\label{eq:H_dipole_lattice},
\end{align}
in the dipole gauge, where $j$ now sums over sites with particle
number $n_{j}=c_{j}^{\dagger}c_{j}$. We have introduced 
\[
\eta\equiv qdf_{0},
\]
as a unitless characterization of the interaction strength between
photons and electrons; $\eta\apprge1$ will correspond to the strong
coupling regime. We are interested in quenched disorder with $V_{j}$
drawn independently on each site from a flat distribution: $V_{j}\in[-W,W]$.
For this lattice model, the gauge transformation becomes $U_{PZW}=\exp\left[-i\sum_{j}qdjn_{j}f_{0}\left(ia^{\dagger}-ia\right)\right]$.
We can use this to return to the Coulomb gauge, for which 
\begin{align}
H_{C} & =\sum_{j}\Bigg[-J\left(e^{-i\eta\left(ia^{\dagger}-ia\right)}c_{j}^{\dagger}c_{j+1}+h.c.\right)+Un_{j}n_{j+1}\nonumber \\
 & \;\;\;\;\;\;+V_{j}n_{j}\Bigg]+\omega a^{\dagger}a\label{eq:H_coulomb_lattice}.
\end{align}
The Eqs. \eqref{eq:H_dipole_lattice} and \eqref{eq:H_coulomb_lattice}
define our main model of interest to be considered throughout this
paper. Note that the gauge choice has no physical effect. For example,
since they are obtained from unitary transforms of one another, $H_{D}$
and $H_{C}$ are isospectral. For numerical simplicity, the majority
of our calculations are done in the dipole gauge, except where stated
otherwise. 

\section{High frequency expansion}

In order to proceed analytically in understanding the dynamics of
the electrons, our goal is to adiabatically eliminate the photons.
In previous work on MBL in cavity QED \cite{MBLCentral1}, we have
done so via a high-frequency expansion, which works when the cavity
frequency $\omega$ is much larger than all other microscopic scales
in the problem. That work implicitly used the dipole gauge and ignored the polarization-squared term due to weak coupling. Here, however, we are interested in the strong coupling, so these approximations are no longer valid. In this section, we will show how a modified high-frequency expansion (HFE) can be obtained which is valid in the strong coupling limit as well as weak coupling.

To start, we note that in the Coulomb gauge there is in fact a reasonable
separation of scales as long as $\omega$ is large. One could proceed
to directly eliminate the cavity photon by a canonical transformation
similar to Schrieffer-Wolff perturbation theory \cite{MBLCentral2}.
However, we instead appeal to a useful trick we developed earlier
in the context of Floquet drive, namely that the Schrieffer-Wolff
transformation can be directly obtained from the effective Hamiltonian
of a high-frequency-driven Floquet system \cite{Bukov2016}. To use this trick, we start by moving to a rotating frame in which the photon
degree of freedom develops time dependence 
\begin{align}
V_\mathrm{rot}(t) & =\exp(i\omega a^{\dagger}at)\nonumber \\
H_\mathrm{rot}(t) & =\sum_{j}\Bigg[-J\left(e^{-i\eta\left(ia^{\dagger}e^{i\omega t}-iae^{-i\omega t}\right)}c_{j}^{\dagger}c_{j+1}+h.c.\right)\nonumber \\
 & \;\;\;\;\;\;+Un_{j}n_{j+1}+V_{j}n_{j}\Bigg]\label{eq:H_rot_floquet}.
\end{align}
Notice that $H_\mathrm{rot}$ is time-periodic with period $T=2\pi/\omega$,
enabling the use of Floquet theory to describe this coupled electron-photon
system. We can then apply Floquet's theorem to rewrite the dynamics
as \cite{HFE1} 
\[
\mathcal{U}_\mathrm{rot}(t)=e^{-iK_\mathrm{eff}(t)}e^{-iH_\mathrm{eff}t}e^{iK_\mathrm{eff}(0)}.
\]
The effective Hamiltonian can then be approximated by a high-frequency
expansion, as detailed in Appendix \ref{sec:hfe_details}. We start
by expanding the Hamiltonian as a Fourier series, $H_\mathrm{rot}(t)=\sum_{\ell}H^{(\ell)}e^{i\ell\omega t}$.
The effective Hamiltonian can be solved perturbatively for large $\omega$
via a high-frequency expansion (HFE) \cite{Goldman2014}: 
\[
H_\mathrm{eff}=H^{(0)}+\sum_{\ell>0}\frac{\left[H^{(\ell)},H^{(-\ell)}\right]}{\ell\omega}+\cdots
\]
Note that we don't actually eliminate the photon from our Hilbert
space. $H_\mathrm{eff}$, as well all of the terms $H^{(\ell)}$, live in
the full Hilbert space $\mathcal{H}=\mathcal{H}_{\mathrm{photon}}\otimes\mathcal{H}_{\mathrm{electrons}}$.
Instead, one can readily show that the photon number commutes with
$H_\mathrm{eff}$ at any order of the HFE. Therefore, we can index the effective
Hamiltonian by the photon number $n$, such that the complicated photon-electron
interactions map to the classical tuning parameter $n$.

At leading (zeroth) order, the effective Hamiltonian is simply given
by the time average: 
\begin{align*}
H_\mathrm{eff}^{0} & =\sum_{j}\left[-J_\mathrm{eff}\left(c_{j}^{\dagger}c_{j+1}+h.c.\right)+Un_{j}n_{j+1}+V_{j}n_{j}\right]\\
J_\mathrm{eff} & =Je^{-\eta^{2}/2}L_{n}\left(\eta^{2}\right),
\end{align*}
where $L_{n}(x)$ is the Laguerre polynomial. The first order correction
leads to correlated hopping of the form $\left(\sum_{j}c_{j}^{\dagger}c_{j+1}\pm h.c.\right)^{2}$,
which are the leading infinite-range integrability-breaking terms
that are generally expected to destroy localization. In the appendix,
we show that this term vanishes in the limit $\omega\to\infty$ at
fixed $\eta$, demonstrating at least asymptotic convergence of the
HFE.

\subsection{Floquet limit ($n\to\infty$)}

In the limit of large photon number, $n\rightarrow\infty$, the effect
of cavity photons becomes identical to that of an external classical
field, i.e., a Floquet drive. We use the asymptotic formula $L_{n}(x/n)\approx\exp\left[x/(2n)\right]\mathcal{J}_{0}(2\sqrt{x})$
to find 
\[
J_\mathrm{eff}(n,\eta)\approx Je^{-\eta^{2}/2}\mathcal{J}_{0}(2\eta\sqrt{n}),
\]
where $\mathcal{J}_{0}$ is the zeroth order Bessel function of the
first kind. This is precisely the formula obtained for Floquet driving
\cite{HFE1}, as expected. It leads to the phenomenon of coherent
destruction of tunneling, whereby tunneling is turned off at leading
order by tuning to one of the zeros of the Bessel function, e.g.,
$2\eta\sqrt{n}\approx2.4$. This idea was used in \cite{CIMBL,CIMBLinherent} to
suggest drive-induced many-body localization since, in the near absence
of tunneling, the disorder should stabilize many-body localization for
sufficiently weak interactions. 

\subsection{Finite photon number}

At finite photon number, the effective hopping depends non-trivially
on both $\eta$ and $n$. In order to have a well-defined Floquet
limit, we clearly need to scale $\eta\sim1/\sqrt{n}$ at large $n$,
but it is less clear what we should do at $n\sim1$. To motivate the
correct scaling, consider the role of weak drive, $\eta\ll1$, for
which we can Taylor expand $J_\mathrm{eff}$: 
\[
\frac{J_\mathrm{eff}}{J}=e^{-\eta^{2}/2}L_{n}(\eta^{2})\approx1-\eta^{2}\left(n+\frac{1}{2}\right).
\]
This suggests to rescale $\eta$ as $1/\sqrt{n+1/2}$. Defining the
rescaled drive strength 
\[
\tilde{\eta}\equiv\eta\sqrt{n+1/2},
\]
we plot $J_\mathrm{eff}(n,\tilde{\eta})$ in Figure \ref{fig:J_eff_hfe}.
At small $\tilde{\eta}$, the different photon numbers match, but
they diverge on a scale $\tilde{\eta}\sim n$. The number of zeros
-- where perfect coherent destruction of tunneling occurs -- is
equal to $n$, as $L_{n}$ is an $n$th order polynomial with positive
real roots. Besides these coherent effects which lead to zeros of
the tunneling, there is an overall Gaussian envelope $\exp\left[-\tilde{\eta}^{2}/\left(2n+1\right)\right]$
which suppresses the tunneling at a low photon number for $\eta\apprge1$.
By contrast, $\eta\to0$ in the limit of large photon number, $n\gg1$,
so we are instead dominated by the oscillatory power-law tails of
the Bessel function.

\begin{figure}[t]
\centering \includegraphics[clip,width=1\linewidth, trim = {0 0 1.5cm 1.5cm}, clip]{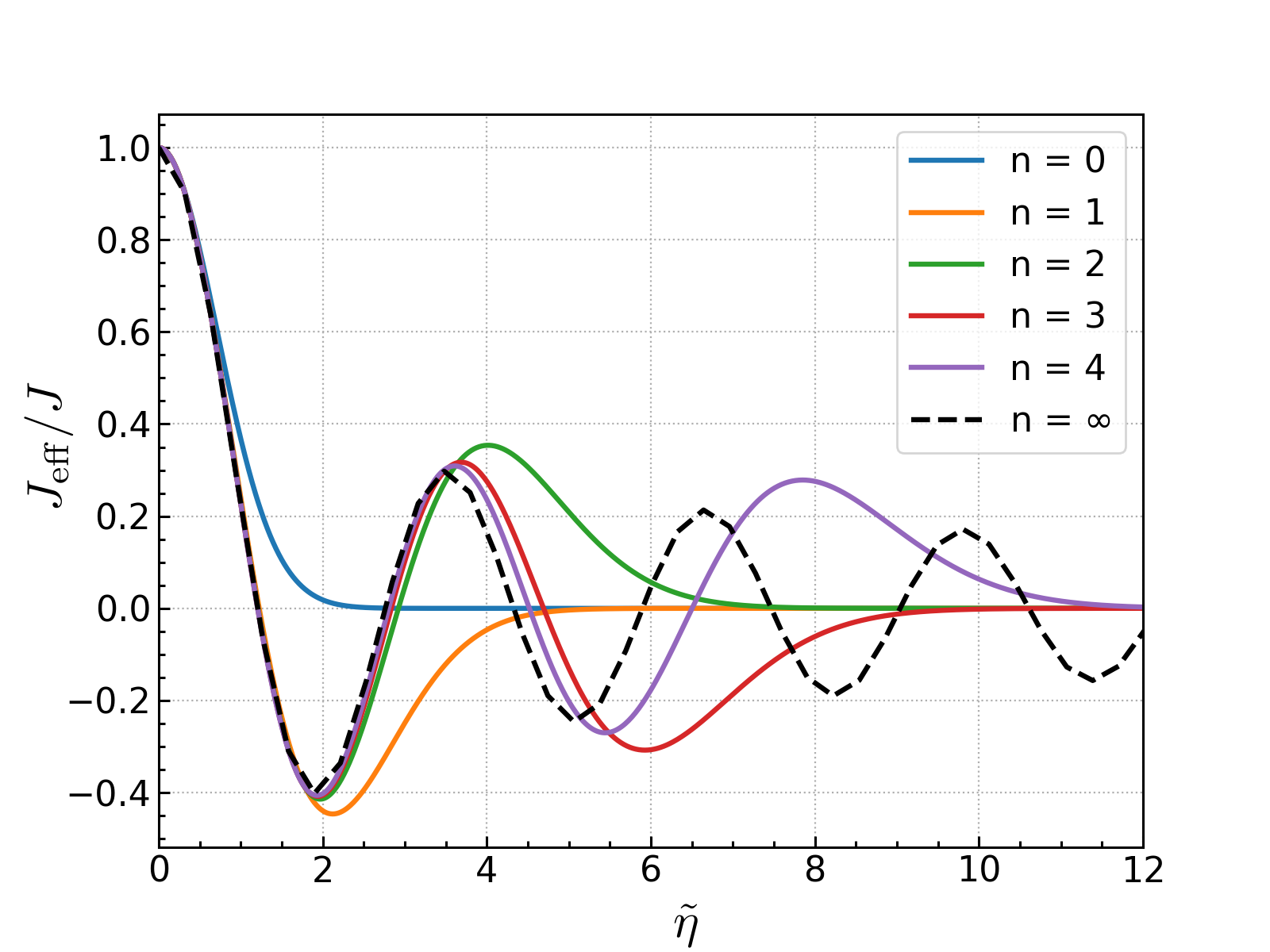}
\caption{Renormalized coupling strength $J_\mathrm{eff}$ as a function of rescaled
drive strength $\tilde{\eta}$ for various photon numbers. The limit
$n\to\infty$ reproduces Floquet drive, where $J_\mathrm{eff}=J\mathcal{J}_{0}(2\tilde{\eta})$.}
\label{fig:J_eff_hfe} 
\end{figure}

The second important effect of finite photon number is an infinite range
correlated hopping, which is captured at leading order by $H_\mathrm{eff}^{1}$.
This term is suppressed as $\omega^{-1}$, as expected for a high-frequency
expansion. More interestingly, it is also suppressed as $1/n$ within the
first order approximation as $n\to\infty$ (see Appendix \ref{sec:hfe_details}), allowing
the definition of a Floquet limit where there is clearly no long-range
interactions. This is similar to the weakly driven case, in which interactions
similarly vanish as $1/(n\omega)$ if the coupling strength is scaled as $1/\sqrt{n}$
(cf. \cite{MBLCentral1}). This suggests that the photon number or coupling
strength at which infinite range correlated hopping will induce delocalization scales
similarly to the weakly coupled case. The main difference is that the scaled coupling
strength $\tilde{\eta}$ has a more interesting dependence, rather than entering
directly at low orders in perturbation theory.

Higher order terms in the HFE will give rise to more complicated interactions,
as well as local dressing of hopping and interactions due to the commutators
of various harmonics with the density-dependent terms in $H_\mathrm{rot}^{(0)}$.
Terms at order $k$ will be suppressed as $1/(n\omega)^{k}$,
so will be negligible in the high frequency and/or large photon number
limit.

We now test predictions of the HFE for two cases: the single particle
regime (Section \ref{sec:single_particle}) and many particle regime
(Section \ref{sec:many_body}).

\section{Single particle localization\label{sec:single_particle}}

If the electron lattice contains a single particle, then the system
is integrable and readily solved numerically. Furthermore, the long-range
correlated hoppings induced by the cavity mode do not play a role,
as terms of the form $c_{j}^{\dagger}c_{j\pm1}c_{j^{\prime}}^{\dagger}c_{j^{\prime}\pm1}$
vanish in the single particle sector unless $j=j^{\prime}\pm1$. Therefore,
higher order terms in the HFE simply become longer range hopping:
second-nearest neighbor at order 1, third-nearest neighbor at order
2, and so on. Combined with modulation of the nearest neighbor hopping
by $J_\mathrm{eff}$ (and higher order corrections), we should simply end
up with a quasilocal hopping model after integrating out the photon.
For our one dimensional model with the chemical potential disorder, such
a system will Anderson localize for any finite disorder strength $W>0$.

\begin{figure*}[t]
\centering \includegraphics[width=1.0\linewidth, trim={0 0 0.5cm 0}, clip]{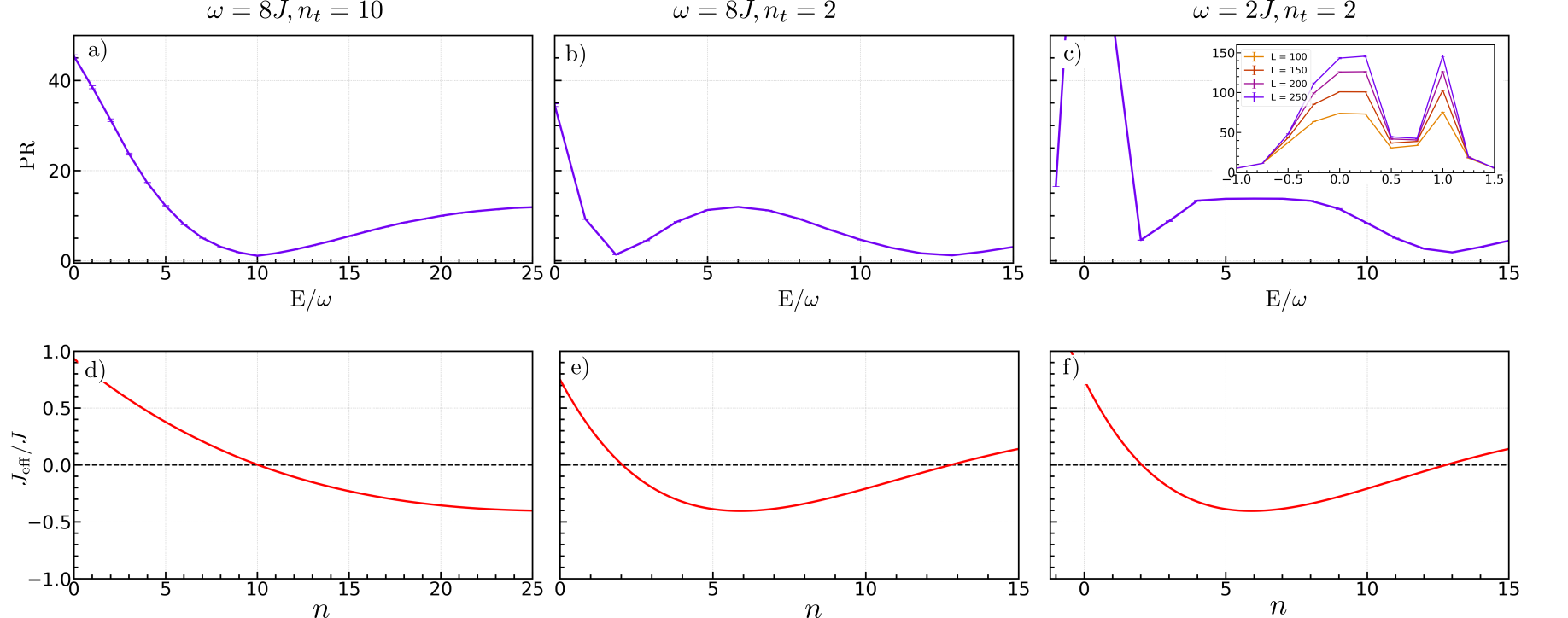}
\caption{(a-c) Participation ratio (PR) as a function for the single-particle model in 3 regimes: (a) $\omega=8J$ and $n_t=10$ ($\eta=0.265$), which approximates the Floquet limit, (b) $\omega=8J$ and $n_t=2$ ($\eta=0.765$), where notable deviations exist from the Floquet limit but the high-frequency expansion (HFE) is still useful, and (c) $\omega=2J$ and $n_t=2$, where the photon frequency is less than the bandwidth and, therefore, the HFE should not hold. For (a-c), disorder is $W=0.5$ and system size is $L=250$. Results are averaged over disorder realizations and energy in the window $E\pm \omega/2$; error bars are smaller than data points. Data is approximately converged to $L=\infty$ except at low energies in (c). The inset to (c) zooms in on this region. (d-f) Effective hopping in the leading order HFE for identical parameters.}
\label{fig:single_ptcl} 
\end{figure*}

However, this is a useful testbed for the controlled analysis of our HFE,
as the microscopic parameters controlling localization depend strongly
on photon number/energy and drive frequency. To test these scalings,
we start by considering single-particle localization as a function
of $\eta$ and energy. We work in the Coulomb gauge, for which the
hopping is translation invariant, and therefore use periodic boundary
conditions to allow translation invariance of our results upon averaging
over disorder. We solve the energy eigenstates of the single particle
model treating the photon exactly up to a cutoff photon number $n_{c}\gg1$.
For all data, including that in the many-body section (Section \ref{sec:many_body}),
we confirm that $n_{c}$ is large enough that our results are unaffected
by it for the range of energies considered. Given that our single
particle model is expected to be localized for all parameters,
we examine the parameter dependence of the localization length to
test how strongly the system localizes. More specifically, we consider
the spatial participation ratio (PR), 
\[
\text{PR}=\left(\sum_{j}p_{j}^{2}\right)^{-1},
\]
where $p_{j}=\langle E|c_{j}^{\dagger}c_{j}|E\rangle$ is the probability
of finding the particle at site $j$ in energy eigenstate $|E\rangle$.
The participation ratio is a useful proxy for the localization length
$\xi$. For a state localized to a single site, $\mathrm{PR}=1$,
while for a delocalized state, $\mathrm{PR}\sim L$.

From the zeroth order HFE, we expect that localization length (and
thus participation ratio) will decrease in line with $J_\mathrm{eff}$, and
in particular will nearly turn off at points where we have coherent
destruction of tunneling. For finite $\omega$, longer range hopping
coming from first order ($\omega^{-1}$) corrections to the HFE should
compete against this. Cavity-modulated hopping in the HFE depends
on three parameters: photon number $n$, cavity frequency $\omega$,
and drive strength $\eta$. In the lab frame, the photon number is not
a conserved quantity. However, it should be closely connected to energy
via $E\approx n\omega$, as all other terms in the Hamiltonian average
to zero. Therefore, we expect that the $n$-dependence in the HFE
should directly reflect in the energy dependence, which is what we
calculate. Note that we choose the convention $E=0$ to be the photon
ground state, rather than the middle of the many-body energy spectrum; 
the latter convention is often used in studies of models with bounded on-site Hilbert
space.

Let's start with the well-understood Floquet limit to understand conventional
coherent destruction of tunneling in our model. We pick a relatively
small disorder $W=0.5$, such that the system has
a large localization length in the absence of photons. Noting from
Figure \ref{fig:J_eff_hfe} that the first zero of $J_\mathrm{eff}$ is
nearly $n$-independent for $n>0$, we choose $\eta$ to target a
photon number $n_{t}$ for which there will first occur coherent destruction
of tunneling. Specifically, we pick $\eta$ such that 
\begin{equation}
2\eta\sqrt{n_{t}+1/2}=x_{0}\implies\mathcal{J}_{0}(2\eta\sqrt{n_{t}+1/2})=0\label{eq:definition_of_n_t},
\end{equation}
where $x_{0}=2.4$ is the first zero of the Bessel function.
Then the Floquet limit is achieved by choosing $n_{t}\gg1$. Data
for $n_{t}=10$ is shown in Figure \ref{fig:single_ptcl}; we further
choose large frequency $\omega=8J$ to suppress higher-order corrections
from the HFE. We see a clear relationship between improved localization
(i.e., smaller PR) and decreased $J_\mathrm{eff}$; in fact, nearly ideal localization with
$\mathrm{PR}=1$ is obtained at $E/\omega =10$. In order to interpolate
$J_\mathrm{eff}$ between integer values of the photon number, we write it
as 
\[
L_{n}(x)\to M(-n,1,x),
\]
where $M$ is the confluent hypergeometric function that is defined
for non-integer $n$.

Next, we deviate away from the Floquet limit by choosing $n_{t}=2$.
In the high-frequency case, $\omega=8J$ (Figure \ref{fig:single_ptcl}),
a similar effect occurs. There are small effects of the quantized
photon being near its vacuum state, but few qualitative changes. The
systems become more interesting when lowering the frequency to allow
the different photon branches to talk to each other. For $\omega=2J$
we see notable effects of the longer range hopping in decreasing
the tendency towards localization. Indeed, the high-frequency
expansion is no longer expected to converge since $\omega$ is smaller than
the bandwidth, $4J$. At low photon number (Figure \ref{fig:single_ptcl}c, inset), 
we see signs of an approach to delocalization. However, the system should 
remain localized, as the low energy states have finite photon numbers and are,
therefore, effectively one-dimensional. 

These results benchmark
the reliability of our HFE, which we now apply to the much more complicated
case of many interacting particles.

\section{Many-body localization\label{sec:many_body}}

In the presence of interactions, localization is far from guaranteed
in one spatial dimension. Indeed, while a vast literature on MBL has
been developed in the past few years (cf. \cite{Nandkishore2015}
for a review), there remain fundamental questions about whether it
even exists as a true phase of matter for infinite times \cite{Suntajs2020,Sels2021,LeBlond2021}.
Setting aside that fundamental point, it is increasingly clear that
small-size numerics miss fundamental properties of MBL, including
giving incorrect critical exponents and likely estimating an incorrect
value of the requisite disorder strength by a significant margin.
These issues will exist in our model as well, in which there is the
additional complication of global (infinite-range) coupling to a cavity
mode.

However, it is clear that cavity-induced coherent destruction of tunneling
will favor localization. In principle, it is probably possible to
localize the system if one tunes $J_\mathrm{eff}$ exactly to zero, as in
that case one simply has commuting density-dependent terms left in
the Hamiltonian. In the presence of the cavity, finite $\omega$ and
infinite range interactions complicate this story.

Here we will explore localization numerically using exact diagonalization,
specifically shift-invert targetting of many-body eigenstates in a
given energy window.
We use two measures to numerically distinguish localized and delocalized
phases of matter. First, we study the energy-dependent level spacing
ratio $r_{n}=\min\{\delta_{n},\delta_{n+1}\}/\max\{\delta_{n},\delta_{n+1}\}$
averaged over different disorder sampling ($\langle r\rangle$), where $\delta_{n}=|E_{n+1}-E_{n}|$
is the gap between neighboring energy levels. For a Poisson distribution,
where the energy levels are randomly distributed, one finds $\langle r\rangle=0.39$.
For delocalized states that satisfy the eigenstate thermalization
hypothesis (ETH \cite{Deutsch1991,Srednicki1994,Rigol2008}), the
level statistics are well-approximated by random matrix theory, and
in particular, undergo level repulsion. Identification of the correct
random matrix ensemble is non-trivial here, but noting that the matrix
elements are all real in the Coulomb gauge and that the energy spectrum
is gauge invariant, we expect to find a Gaussian orthogonal ensemble
(GOE) statistics, for which $\langle r\rangle=0.53$ \cite{LSR1,LSR2}.

Second, we calculate the classical relative entropy of neighboring energy eigenstates, also known as the Kullback-Liebler divergence (KL), $KL=-\sum_{i}p_{i,n}\ln\frac{p_{i,n}}{p_{i,n+1}}$,
where $p_{i,n}$ is the probability to find the state at energy $E_{n}$
in basis state $i$. We use the natural direct product basis $i$ consisting of fermions localized
in real space at half-filling and photons in their Fock states.
It has been shown recently that the scaling
behavior of KL with system size gives strong evidence for localization/delocalization
\cite{KL1,KL2}. In particular, $KL=2$ for delocalized states, while
$KL\propto L$ when the system is localized.

We calculate these metrics for localization at half filling and for
two different target photon numbers: $n_{t}=2$ and $n_{t}=6$. For historical reasons, calculations of $r$ and $KL$ are done with the 
opposite sign of $J$ ($J\to-J$) and modified cavity interactions
corresponding to a staggering chemical potential in the dipole gauge: 
$\sum_{j}jn_{j}\to\sum_{j}(-1)^{j}n_{j}/2$.
As shown in Appendix \ref{sec:hfe_details}, this leaves $H_\mathrm{eff}^{0}$
unchanged, while making minor changes to higher order corrections.
Therefore, we do not anticipate that this choice will significantly
affect the results.

The results, shown in Figure \ref{fig:mbl_full_ed}, indicate that,
indeed, the system pushes towards localization when $J_\mathrm{eff}\to0$
in the leading-order HFE. Localization is noticeably weaker for smaller
$n_{t}=2$ compared to larger $n_{t}=6$ which approaches the Floquet
limit. One possible explanation for this effect is that the energy
eigenstates are not fully localized in photon space, despite the predictions
of the HFE, due to hybridization between photon levels at large system
size $L$. This can spread the photon state over a range of $J_\mathrm{eff}$,
many of which will of course not be zero, and therefore hinder localization.
The effect should be stronger at small $n_{t}$ because relative fluctuations
of the photon (and thus $J_\mathrm{eff}$) are larger there.

\begin{figure}[t]
\centering \includegraphics[trim=0.8cm 1.4cm 1.0cm .2cm, clip=true,width=1\columnwidth]{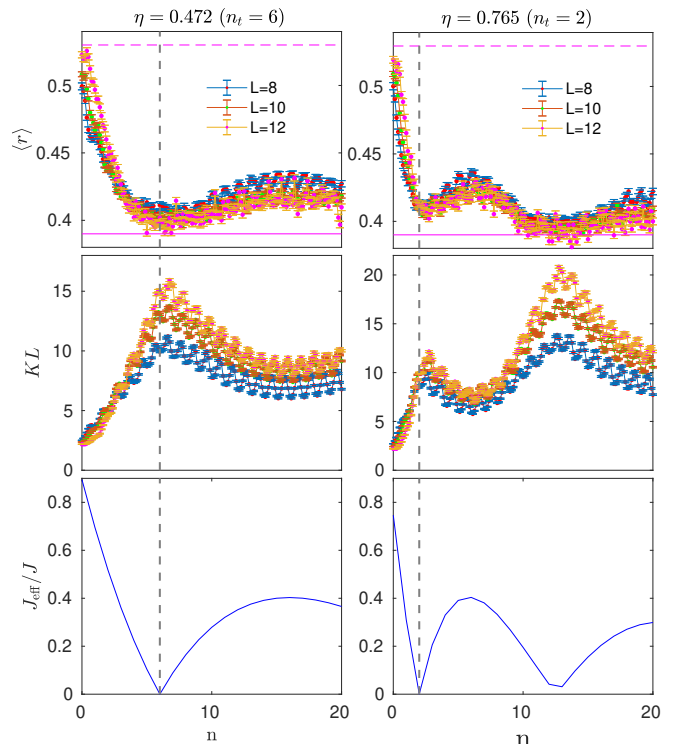}\caption{Level statistic $r$ (top) and Kullback-Leibler divergence (KL) (middle)
for $n_{t}=6$ (left) and $n_{t}=2$ (right) as a function of energy,
compared with hopping strength $J_\mathrm{eff}$ obtained from the HFE (bottom).
Both large and small target photon number show evidence of localization
when energy is chosen to get coherent destruction of tunneling, $J_\mathrm{eff}\approx0$.
Localization is stronger for $n_{t}=6$, which we attribute to a smaller
relative fluctuation of the photon number at higher energy. Note that the model
has been slightly modified compared to other figures, namely $J\to -J$ and $\sum_{j}jn_{j}\to\sum_{j}(-1)^{j}n_{j}/2$,
as described in the text.}
\label{fig:mbl_full_ed} 
\end{figure}

While the regime of MBL or, at least, significant slowing down of
the dynamics remain only loosely clear, what is clear is that localization
is favored by small $J_\mathrm{eff}$ at high energy. This model therefore
illustrates a variant of the inverted mobility edge that we found
in an earlier work \cite{MBLCentral2}, namely that such centrally
coupled models tend to localize at high energies and delocalize at
low energies. The coherent oscillations of $J_\mathrm{eff}(n)$ lead to an
even more interesting structure where the thermal phase can occur
in multiple windows of energy, for which $J_\mathrm{eff}$ is above the threshold
for thermalization.

Another interesting note is that there are large regions of the data
where both level statistics and KL divergence seem to stay in between
the MBL and ETH values with little dependence on system size. This
is likely a finite size effect that will eventually flow to one phase
or the other (likely the thermal phase, which is stabilized by higher-order
corrections and resonances that are beyond our small size numerics).
However, this model seems to have surprisingly slow finite size drifts
compared to similar models in the MBL literature. 

Finally, to emphasize the relevance of being able to tune localization
dynamics via energy, we consider a more physically relevant observable,
namely the imbalance $\mathcal{I}$, which in the absence of a cavity
has been measured in multiple AMO experiments on MBL \cite{landig2016quantum, Hruby2018avalanche, Mazzucchi:16, PhysRevA.94.023632}. The idea is to initialize
a charge density wave (CDW) with minimal wavelength, $|0101\cdots\rangle$,
and then watch for relaxation of the CDW order as the system time
evolves. The imbalance is defined as 
\[
\mathcal{I}=L^{-1}\sum_{j}\left(-1\right)^{j}\left(2\langle n_{j}\rangle-1\right),
\]
which measures the order such that $\mathcal{I}=1$ in the initial
state and $\mathcal{I}=0$ in thermal equilibrium. In order to modify
the energy, we prepare the cavity mode in a coherent state $|\alpha\rangle$,
such that the disorder-averaged energy $\overline{E}$ is 
\begin{align*}
\overline{E}=&\overline{\langle\alpha,0101\cdots|H_{D}|\alpha,0101\cdots\rangle}
\\=&\overline{\sum_{j\text{ odd}}V_{j}}+\omega\left|\alpha\right|^{2}=\omega\left|\alpha\right|^{2},
\end{align*}
Note that we have chosen an initial state with zero polarization,
$\sum_{j}j\langle n_{j}\rangle=0$, such that switching from Coulomb
gauge to dipole gauge does not change the initial state \footnote{In practice we do this by choosing to define $j=0$ such that polarization
vanishes, even if that means the origin lies between sites. This choice
has no effect on the physics, but makes the calculation easier.}. Then we measure $\langle\mathcal{I}(t)\rangle$, the results of
which are plotted in Figure \ref{fig:imbalance} for 300-6000 realization of disorders. We see strong indications
of slowing down and localization when tuning near the point with $J_\mathrm{eff}=0$.
While this does not prove the existence of an MBL phase, it does suggests
an experimentally relevant route to tuning localization dynamics in
these cavity QED systems.

\begin{figure}[t]
\centering \includegraphics[clip,width=0.5\textwidth]{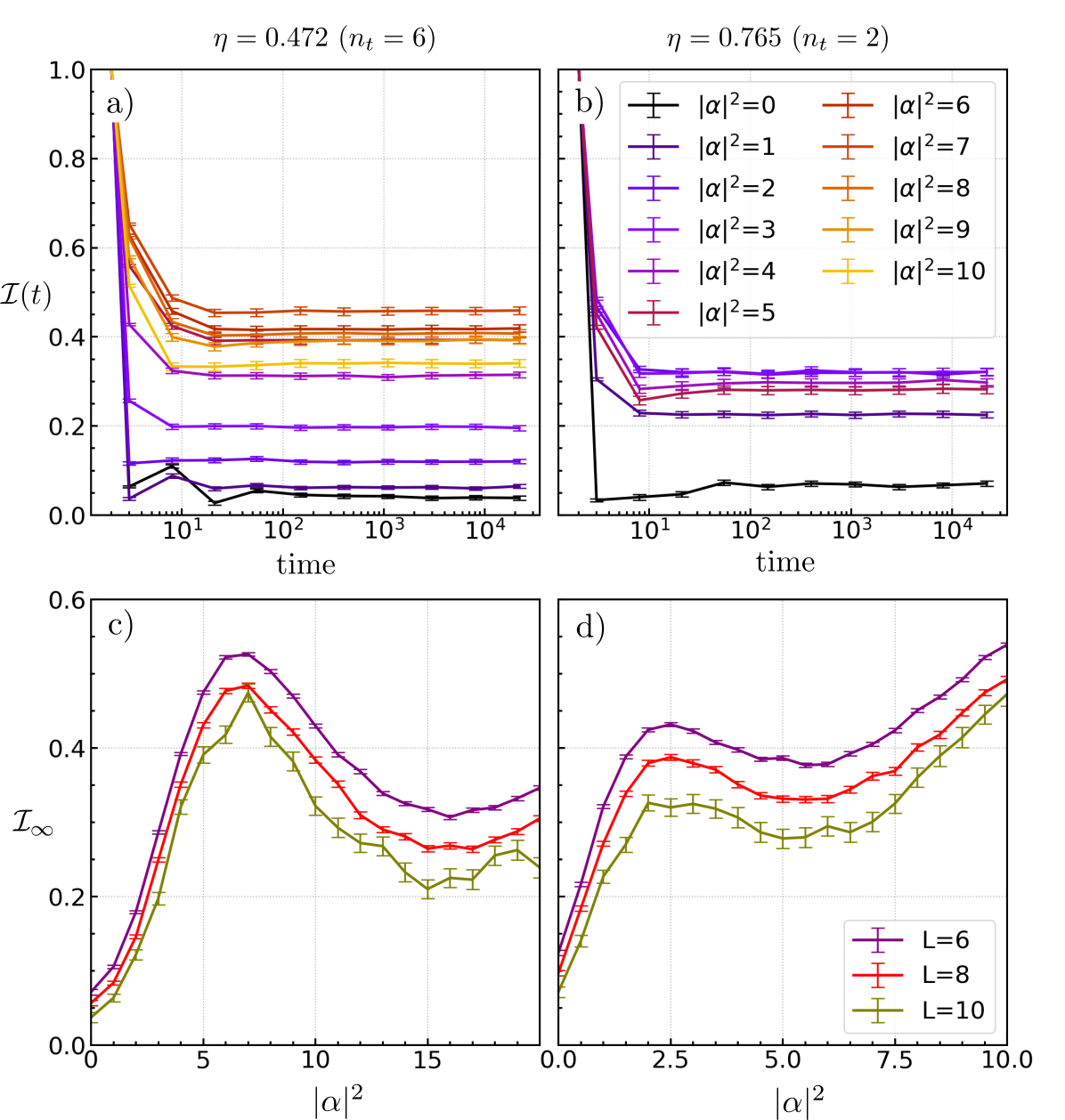}\caption{(a and b) Imbalance ($\mathcal{I}(t)$) as a function of time for different $|\alpha|^2$ and $L=10$. (c and d) Steady state value, $I(t=\infty)$. For small finite size, $\mathcal{I}(\infty)$ is nonzero, with a value that aligns well with other metrics for localization in Figure \ref{fig:mbl_full_ed}. For values with $J_\mathrm{eff}=0$ from the HFE, $\mathcal{I}(\infty)$ is large and decreases slowly with system size, consistent with many-body localized dynamics. Parameters are $\omega=8J$, $W=0.5J$ and $U=1.5J$ with constant electric field ($\eta_{j}=\eta=\mathrm{const.}$) as in the single-particle case. The initial state is a period-2 charge density wave with photons in coherent state $|\alpha\rangle$.}
\label{fig:imbalance} 
\end{figure}

\section{Experimental realization}

As alluded to in Section \ref{sec:setup}, the main experimental setup
where this will be relevant is electronic systems in the strong coupling
regime of cavity QED. Semiconducting nanowires are a natural situation
to consider these physics, as illustrated in Figure \ref{fig:setup}.
The one-dimensional nature of this model is not crucial for realizing
cavity-mediated tuning of the hopping. In higher dimensional models,
a similar analysis will hold, with coupling strength $\eta_{ij}$
for a given bond $<ij>$ proportional to the component of the electric
field along the direction of the hopping. In two dimensions, for example,
this can be used to tune lattice symmetries and connectivities, as
we recently showed the case of classical drive on a moir\'e heterostructure
\cite{GeArxiv2021}. We also note that the strong coupling regime
has been achieved in electron systems, most notably cavity-mediated
chemical reactions \cite{ribeiro2018polariton, schafer2021shining},
which is precisely the system for which gauge choices in cavity QED
have been most actively explored \cite{gauge3,gauge2,gauge1,gauge}.

Modifying the hopping strength will have a profound impact on the
properties of these electronic systems. While molecular Hamiltonians
are much more complicated than those we have written down due to strong
electron-electron interactions, cavity-mediated control of the kinetic
energy will still significantly modify their electron structure and
many allow control of chemical reactions via tuning the strength of
a cavity-pumping laser to change the intracavity photon number. In
the context of nanowires, laser tuning of the photon state will directly
modify transport across the wire. By gating the nanowire into a low-filling
regime, the single-particle physics from Section \ref{sec:single_particle}
may be accessible. There we expect exponential dependence of conductivity
with $J_\mathrm{eff}$, as the transport in such single-particle localized
systems is primarily through thermally activated variable range hopping
\cite{Mott1969,Efros1975}. In both the cases of quantum chemistry
and nanowires, the cavity plays an important role in enhancing the
electric field strength at the electrons, allowing a strong drive
regime to be accessed. A key result of our work is that the side effect
of the cavity -- photon-mediated long-range coupling -- does not
immediately destroy localization.

Cavity-induced localization may also be relevant to other cavity QED
settings. It's not likely to be immediately accessible in the context
of neutral atoms, as these systems are difficult to push into the
strong coupling regime. However, cavity QED with superconducting qubits,
a.k.a. circuit QED, could be a useful platform to realize these physics.
In that case the dipole moment of the artificial atoms, i.e., the
superconducting qubits, are local to each qubit and there will not
be direct hopping of the electronic excitations. Instead, the spatially
dependent coupling (in the dipole gauge) could be realized by the
mode profile $f(x_{j})$ of the superconducting cavity, where $x_{j}$
are the spatial locations of the superconducting qubits along the
cavity. Locally interacting Bose-Hubbard models with tunable frequencies
or disorder are readily realized in recent transmon experiments \cite{Chen2014,Roushan2014,Neill2016,Roushan2017,Roushan2017a}.
Placing an array of transmons at chosen locations along the superconducting
cavity in order to realize an effective potential gradient, $f(x_{j})\propto j$,
and tuning into the strong coupling regime, similar cavity-mediated
localization should be realizable. In these controllable quantum systems,
local measurements such as the imbalance $\mathcal{I}$ are natural,
and would give a strong indication of localization dynamics.

\section{conclusion}
In conclusion, we studied the possibility of realizing many-body localization
(MBL) in the context of strong coupling cavity QED. By working with
a fermion Hubbard chain inside a high-quality cavity, we show that
a variant of coherent destruction of tunneling is possible even at
a moderate photon number. In the single-particle regime, this directly
modulates the localization length. In the many-body regime, we show
evidence that MBL survives, despite the presence of cavity-mediated
long-range correlated hopping and interactions. These results should
prove relevant to electronic systems coupled to optical cavities,
which routinely access the strong coupling regime \cite{gauge3,gauge2,gauge1,gauge},
and may also be relevant to superconducting circuit QED.

A major theoretical accomplishment of this work is to develop a method
for integrating out the cavity photon in the presence of strong driving,
which is intricately tied to the electromagnetic gauge choice made
in quantizing the cavity mode. The high-frequency expansion developed
here can be extended to other cases of strong and/or resonant drive,
where a naive expansion in the dipole gauge would fail. A particularly interesting case where we anticipate using these results is the cavity
QED realization of a quantum time crystal, in which the cavity mode
plays the role of a Floquet drive. Since both the time crystal and
most other non-trivial Floquet phases of matter require strong and/or
resonant drive, these techniques will be valuable in understanding
how these phases behave in the presence of quantized photons.

\section{Acknowledgments}

The authors acknowledge useful discussions with Nathan Ng. Work was
performed with support from the National Science Foundation through
award number DMR-1945529 (MK and SRK) and the Welch Foundation through
award number AT-2036-20200401 (MK and RG). Part of this work was performed
at the Aspen Center for Physics, which is supported by National Science
Foundation grant PHY-1607611, and at the Kavli Institute for Theoretical
Physics, which is supported by the National Science Foundation under
Grant No. NSF PHY1748958. RG is also supported by the "Fundamental Research Funds for the Central Universities". We used the computational resources of the
Lonestar 5 cluster operated by the Texas Advanced Computing Center
at the University of Texas at Austin and the Ganymede and Topo clusters
operated by the University of Texas at Dallas' Cyberinfrastructure
and Research Services Department.   


\begin{thebibliography}{[1]}
\bibitem{}
In practice we do this by choosing to define j = 0 such that polarization vanishes, even if that means the origin lies between
sites. This choice has no effect on the physics, but makes the calculation easier.
\end{thebibliography}

 \newpage{}

\appendix
\onecolumngrid

\section{Details of high-frequency expansion\label{sec:hfe_details}}

We start the high-frequency expansion (HFE) using the Floquet version
of our Hamiltonian, Eq. \eqref{eq:H_rot_floquet}: 
\[
H_\mathrm{rot}(t)=\sum_{j}\left[-J\left(e^{-i\eta_{j}\left(ia^{\dagger}e^{i\omega t}-iae^{-i\omega t}\right)}c_{j}^{\dagger}c_{j+1}+h.c.\right)+Un_{j}n_{j+1}+V_{j}n_{j}\right]
\]
Note that we allow position-dependent coupling $\eta_{j}$. This can
capture our two main cases from the text: constant electric field,
$\eta_{j}=\eta$, and staggered chemical potential, $\eta_{j}=\eta\left(-1\right)^{j}$.
We recognize the photon-dependent term as the displacement operator,
\[
\mathcal{D}(\alpha)=\exp\left[\alpha a^{\dagger}-\alpha^{\ast}a\right]
\]
where $\alpha=\eta_{j}e^{i\omega t}$. The Hermitian conjugate ($+h.c.$)
is then 
\[
e^{i\eta_{j}\left(-iae^{-i\omega t}+ia^{\dagger}e^{i\omega t}\right)}c_{j+1}^{\dagger}c_{j}=\mathcal{D}(-\alpha)c_{j+1}^{\dagger}c_{j}
\]
where we used that $\eta_{j}$ is real. In order to do a high frequency
expansion, we must expand $H_\mathrm{rot}$ in Fourier harmonics: 
\[
H_\mathrm{rot}(t)=\sum_{\ell}H_\mathrm{rot}^{(\ell)}e^{i\ell\Omega t}.
\]
Let's start by doing so for the displacement operator, from which
the remainder will follow. First, note that the matrix elements of
$\mathcal{D}$ in the number basis are given by \cite{Cahill1969}
\begin{align*}
\langle m|\mathcal{D}(\alpha)|n\rangle & =\begin{cases}
\sqrt{\frac{n!}{m!}}\alpha^{m-n}e^{-\left|\alpha\right|^{2}/2}L_{n}^{(m-n)}\left(\left|\alpha\right|^{2}\right) & \text{for }m\geq n\\
\sqrt{\frac{m!}{n!}}\left(-\alpha\right)^{n-m}e^{-\left|\alpha\right|^{2}/2}L_{m}^{(n-m)}\left(\left|\alpha\right|^{2}\right) & \text{for }m<n
\end{cases}
\end{align*}
where $L_{n}^{(k)}$ are the associate Laguerre polynomials. The only
time-dependence here is in the $\alpha^{m-n}$ or $\left(-\alpha\right)^{m-n}$
term. Noting that $\alpha^{m-n}=\eta_{j}^{m-n}e^{i\omega\left(m-n\right)t}$,
this is clearly going to give the $\ell$th Fourier harmonic with
$\ell=m-n\geq0$. Adding together all the terms, the $\ell$th Fourier
harmonic of $H_\mathrm{rot}$ is then 
\begin{align*}
H_\mathrm{rot}^{(\ell\ge0)} & =\sum_{n=0}^{\infty}\sum_{j}\Bigg[-J\left(\sqrt{\frac{n!}{\left(n+\ell\right)!}}\eta_{j}^{\ell}e^{-\left|\eta_{j}\right|^{2}/2}L_{n}^{(\ell)}\left(\left|\eta_{j}\right|^{2}\right)c_{j}^{\dagger}c_{j+1}+\sqrt{\frac{n!}{\left(n+\ell\right)!}}\left(-\eta_{j}\right)^{\ell}e^{-\left|\eta_{j}\right|^{2}/2}L_{n}^{(\ell)}\left(\left|\eta_{j}\right|^{2}\right)c_{j+1}^{\dagger}c_{j}\right)\\
 & \;\;\;\;\;\;\;\;\;\;\;\;+\delta_{\ell,0}\left(Un_{j}n_{j+1}+V_{j}n_{j}\right)\Bigg]|n+\ell\rangle\langle n|\\
 & =\sum_{n=0}^{\infty}\sum_{j}\Bigg[-J\sqrt{\frac{n!}{\left(n+\ell\right)!}}\eta_{j}^{\ell}e^{-\left|\eta_{j}\right|^{2}/2}L_{n}^{(\ell)}\left(\left|\eta_{j}\right|^{2}\right)\left(c_{j}^{\dagger}c_{j+1}+\left(-1\right)^{\ell}c_{j+1}^{\dagger}c_{j}\right)+\delta_{\ell,0}\left(Un_{j}n_{j+1}+V_{j}n_{j}\right)\Bigg]|n+\ell\rangle\langle n|\\
H_\mathrm{rot}^{(-\ell)} & =\left[H_\mathrm{rot}^{(\ell)}\right]^{\dagger}
\end{align*}
where the last line follows from Hermitianity of $H_\mathrm{rot}$.

The time-average of $H_\mathrm{rot}$, which is the zeroth order $H_\mathrm{eff}$,
is then easy to compute. Noting that the effective Hamiltonian has
zero frequency and is, therefore, diagonal in the number basis, we
can project onto the photon number state $|n\rangle$ to get 
\begin{align*}
H_\mathrm{eff}^{0}(n) & =\langle n|H_\mathrm{rot}^{(0)}|n\rangle\\
 & =\sum_{j}\Bigg[-\underbrace{Je^{-\left|\eta_{j}\right|^{2}/2}L_{n}\left(\left|\eta_{j}\right|^{2}\right)}_{J_\mathrm{eff}(n,\eta_{j})}\left(c_{j}^{\dagger}c_{j+1}+c_{j+1}^{\dagger}c_{j}\right)+Un_{j}n_{j+1}+V_{j}n_{j}\Bigg]
\end{align*}
Note that, for both choices of $\eta_{j}$ made in paper, one gets
an identical $J_\mathrm{eff}$ at zeroth order.

The first order correct will give rise to leading correlated hopping
terms. Again, projecting onto the $n$ photon state, we have 
\begin{align*}
H_\mathrm{eff}^{1}(n) & =\sum_{\ell>0}\frac{\langle n|\left[H^{(\ell)},H^{(-\ell)}\right]|n\rangle}{\ell\omega}=\sum_{\ell>0}\frac{\langle n|\left[H^{(\ell)},\left(H^{(\ell)}\right)^{\dagger}\right]|n\rangle}{\ell\omega}\\
 & =\sum_{\ell>0}\frac{\langle n|H^{(\ell)}\left(H^{(\ell)}\right)^{\dagger}|n\rangle-\langle n|\left(H^{(\ell)}\right)^{\dagger}H^{(\ell)}|n\rangle}{\ell\omega}\\
 & =\sum_{\ell=1}^{n}\frac{\langle n|H^{(\ell)}|n-\ell\rangle\langle n-\ell|\left(H^{(\ell)}\right)^{\dagger}|n\rangle}{\ell\omega}-\sum_{\ell=1}^{\infty}\frac{\langle n|\left(H^{(\ell)}\right)^{\dagger}|n+\ell\rangle\langle n+\ell|H^{(\ell)}|n\rangle}{\ell\omega}\\
 & =\sum_{\ell=1}^{n}\frac{\langle n|H^{(\ell)}|n-\ell\rangle\left(\langle n|H^{(\ell)}|n-\ell\rangle\right)^{\dagger}}{\ell\omega}-\sum_{\ell=1}^{\infty}\frac{\left(\langle n+\ell|H^{(\ell)}|n\rangle\right)^{\dagger}\langle n+\ell|H^{(\ell)}|n\rangle}{\ell\omega}\\
 & =\frac{J^{2}}{\omega}\sum_{\ell=1}^{n}\frac{\left(n-\ell\right)!}{\ell n!}\sum_{j,j^{\prime}}L_{n-\ell}^{(\ell)}\left(\left|\eta_{j}\right|^{2}\right)L_{n-\ell}^{(\ell)}\left(\left|\eta_{j^{\prime}}\right|^{2}\right)\eta_{j}^{\ell}\eta_{j^{\prime}}^{\ell}e^{-\left(\left|\eta_{j}\right|^{2}+\left|\eta_{j^{\prime}}\right|^{2}\right)/2}\times\\
 & \;\;\;\;\;\;\;\;\;\;\;\;\;\;\;\;\times\left(-1\right)^{\ell}\left(c_{j}^{\dagger}c_{j+1}+\left(-1\right)^{\ell}c_{j+1}^{\dagger}c_{j}\right)\left(c_{j^{\prime}}^{\dagger}c_{j^{\prime}+1}+\left(-1\right)^{\ell}c_{j^{\prime}+1}^{\dagger}c_{j^{\prime}}\right)\\
 & \;\;\;\;\;-\frac{J^{2}}{\omega}\sum_{\ell=1}^{\infty}\frac{n!}{\ell\left(n+\ell\right)!}\sum_{j,j^{\prime}}L_{n}^{(\ell)}\left(\left|\eta_{j}\right|^{2}\right)L_{n}^{(\ell)}\left(\left|\eta_{j^{\prime}}\right|^{2}\right)\eta_{j}^{\ell}\eta_{j^{\prime}}^{\ell}e^{-\left(\left|\eta_{j}\right|^{2}+\left|\eta_{j^{\prime}}\right|^{2}\right)/2}\times\\
 & \;\;\;\;\;\;\;\;\;\;\;\;\;\;\;\;\times\left(-1\right)^{\ell}\left(c_{j}^{\dagger}c_{j+1}+\left(-1\right)^{\ell}c_{j+1}^{\dagger}c_{j}\right)\left(c_{j^{\prime}}^{\dagger}c_{j^{\prime}+1}+\left(-1\right)^{\ell}c_{j^{\prime}+1}^{\dagger}c_{j^{\prime}}\right)
\end{align*}
In order to simplify this expression, let's consider the simplest
case that $\eta_{j}=\eta$ is constant. Then 
\begin{equation}\label{Supp:HeffOrder1}
\begin{split}
H_\mathrm{eff}^{1}(n) & =\frac{J^{2}}{\omega}\sum_{\ell=1}^{n}\frac{\left(-1\right)^{\ell}}{\ell}\left[\frac{\left(n-\ell\right)!}{n!}\left(L_{n-\ell}^{(\ell)}\left(\eta^{2}\right)\right)^{2}-\frac{n!}{\left(n+\ell\right)!}\left(L_{n}^{(\ell)}\left(\eta^{2}\right)\right)^{2}\right]\eta^{2\ell}e^{-\eta^{2}}\times\\
 & \;\;\;\;\;\;\;\;\;\;\;\;\;\;\;\;\times\sum_{j,j^{\prime}}\left(c_{j}^{\dagger}c_{j+1}+\left(-1\right)^{\ell}c_{j+1}^{\dagger}c_{j}\right)\left(c_{j^{\prime}}^{\dagger}c_{j^{\prime}+1}+\left(-1\right)^{\ell}c_{j^{\prime}+1}^{\dagger}c_{j^{\prime}}\right)\\
 & \;\;\;\;\;-\frac{J^{2}}{\omega}\sum_{\ell=n}^{\infty}\frac{\left(-1\right)^{\ell}n!}{\ell\left(n+\ell\right)!}\left(L_{n}^{(\ell)}\left(\eta^{2}\right)\right)^{2}\eta^{2\ell}e^{-\eta^{2}}\sum_{j,j^{\prime}}\left(c_{j}^{\dagger}c_{j+1}+\left(-1\right)^{\ell}c_{j+1}^{\dagger}c_{j}\right)\left(c_{j^{\prime}}^{\dagger}c_{j^{\prime}+1}+\left(-1\right)^{\ell}c_{j^{\prime}+1}^{\dagger}c_{j^{\prime}}\right)\\
 & =-\frac{J^{2}e^{-\eta^{2}}}{\omega}\left(\sum_{\ell=1,3,\cdots}\frac{f(n,\ell,\eta)}{\ell}\eta^{2\ell}\right)\left[\sum_{j}\left(c_{j}^{\dagger}c_{j+1}-c_{j+1}^{\dagger}c_{j}\right)\right]^{2}\\
 & \;\;\;+\frac{J^{2}e^{-\eta^{2}}}{\omega}\left(\sum_{\ell=2,4,\cdots}\frac{f(n,\ell,\eta)}{\ell}\eta^{2\ell}\right)\left[\sum_{j}\left(c_{j}^{\dagger}c_{j+1}+c_{j+1}^{\dagger}c_{j}\right)\right]^{2}
 \end{split}
\end{equation}
where 
\[
f(n,\ell,\eta)=\begin{cases}
\frac{\left(n-\ell\right)!}{n!}\left(L_{n-\ell}^{(\ell)}\left(\eta^{2}\right)\right)^{2}-\frac{n!}{\left(n+\ell\right)!}\left(L_{n}^{(\ell)}\left(\eta^{2}\right)\right)^{2} & \text{if }\ell\leq n\\
-\frac{n!}{\left(n+\ell\right)!}\left(L_{n}^{(\ell)}\left(\eta^{2}\right)\right)^{2} & \text{if }\ell>n
\end{cases}
\]
The case $\eta_{j}=\left(-1\right)^{j}$ is quite similar, except
with an additional distinction between hopping sign for even and odd
sites. From Eq.~\ref{Supp:HeffOrder1}, we can define the effective long-range coupling as
\begin{align}
\label{geffMainbody}
    g_\mathrm{eff}^\mathrm{I} & = e^{-\eta^{2}}\sum_{\ell=1,3,\cdots}\frac{f(n,\ell,\eta)}{\ell}\eta^{2\ell}\\
    g_\mathrm{eff}^\mathrm{R} & = e^{-\eta^{2}} \sum_{\ell=2,4,\cdots}\frac{f(n,\ell,\eta)}{\ell}\eta^{2\ell}
\end{align}
where the labels R and I denote that the corresponding correlated hopping terms have real and imaginary phase, respectively.
The behavior of these $g_\mathrm{eff}$ couplings as a function of $\eta$ is shown in Figure \ref{fig:g_eff_hfe}. For any $n$, the coupling constants clearly vanish at large $\tilde{\eta}$, meaning that strong coupling eliminates long-range correlated hopping as well as nearest neighbor hopping. Combined with our arguments for the appropriate scaling of $g$ with $L$ (see main text), this provides further analytical support for the possibility of MBL in the strong coupling regime, even for the extreme quantum limit of photons in their ground state ($n=0$). Note, however, that at medium coupling strengths $\tilde \eta \sim 1$, $g_\mathrm{eff}$ for the real and imaginary terms do not vanish at the same values of $\tilde \eta$, implying that the fine-tuned physics of coherent destruction of tunneling do not simply persist in these higher order terms.

\begin{figure}[t]
\centering \includegraphics[clip,width=1\linewidth, trim = {0 0 1.5cm 0.5cm}, clip]{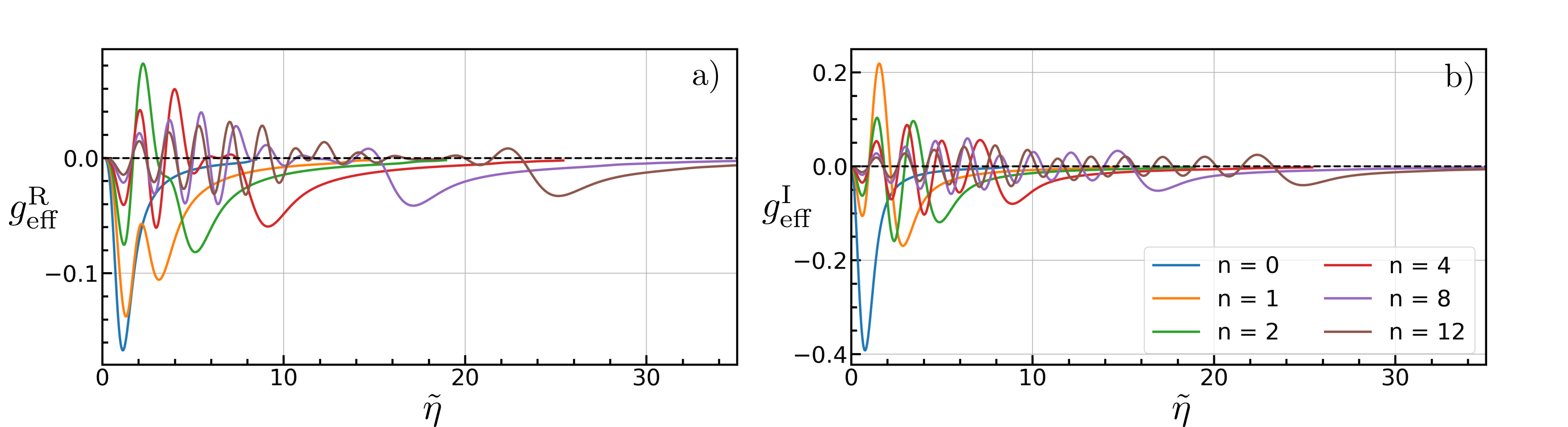}
\caption{Effective long-range hopping coefficient $g_\mathrm{eff}$ as a function of rescaled
drive strength $\tilde{\eta}$ for various photon numbers. At large drive strength, $\tilde \eta \gg 1$, the long-range correlated hopping is switched off, favoring localization.}
\label{fig:g_eff_hfe} 
\end{figure}

We are often interested in the large-$n$ expansion of this effective long-range
coupling, since that will tell us about the semi-classical limit of the cavity photons. 
Noting that $\eta=\tilde{\eta}/\sqrt{n+1/2}$, a large $n$ expansion
corresponds to $\eta\ll1$. Thus, at leading order, we simply have
$e^{-\eta^{2}}\approx1$. Then the relevant term is $f(n,\ell,\eta)\eta^{2\ell}$.
We will see this is dominated by small $\ell$, so for now restrict
ourselves to $\ell\ll n$. Writing out the associated
Legendre polynomial,
\begin{equation}
L_{n}^{(\ell)}\left(\eta^{2}\right)=\sum_{m=0}^{n}\frac{\left(-\eta^{2}\right)^{m}\left(n+\ell\right)!}{\left(n-m\right)!m!\left(\ell+m\right)!}
\end{equation}
Taking this at second order in $\eta\ll1$ (i.e., up to $m=2$), we
have
\begin{align}
L_{n}^{(\ell)}\left(\eta^{2}\right) & =\frac{\left(n+\ell\right)!}{n!\ell!}-\eta^{2}\frac{\left(n+\ell\right)!}{\left(n-1\right)!\left(\ell+1\right)!}+\frac{\eta^{4}}{2}\frac{\left(n+\ell\right)!}{\left(n-2\right)!\left(\ell+2\right)!}+O\left(\eta^{6}\right)\\
 & =\frac{\left(n+\ell\right)!}{n!\ell!}\left[1-\frac{\eta^{2}n}{\ell+1}+\frac{\eta^{4}n\left(n-1\right)}{\left(\ell+1\right)\left(\ell+2\right)}+O\left(\eta^{6}\right)\right]\\
\left[L_{n}^{(\ell)}\left(\eta^{2}\right)\right]^{2} & =\left(\frac{\left(n+\ell\right)!}{n!\ell!}\right)^{2}\left[1-\eta^{2}\frac{2n}{\ell+1}+\eta^{4}\frac{2n\left(n-1\right)\left(\ell+1\right)+n^{2}\left(\ell+2\right)}{\left(\ell+1\right)^{2}\left(\ell+2\right)}+O\left(\eta^{6}\right)\right]
\end{align}
Notice that each term in the square brackets is independent of $n$
in the limit $n\gg\ell$, since $\eta^{2}\approx\tilde{\eta}^{2}/n$.
Furthermore, higher order terms $\sim\eta^{2m}$ quickly vanish for
$m\gg1$ due to the factorial appearing in the denominator. 

When we combine terms to get $f$, we see that (for $\ell\ll n$)
\begin{align}
f(n,\ell,\eta) & =\frac{n!}{\left(n-\ell\right)!\left(\ell!\right)^{2}}\left[1-\eta^{2}\frac{2\left(n-\ell\right)}{\ell+1}+\eta^{4}\frac{2\left(n-\ell\right)\left(n-\ell-1\right)\left(\ell+1\right)+\left(n-\ell\right)^{2}\left(\ell+2\right)}{\left(\ell+1\right)^{2}\left(\ell+2\right)}+O\left(\eta^{6}\right)\right]\\
 & \;\;\;\;-\frac{\left(n+\ell\right)!}{n!\left(\ell!\right)^{2}}\left[1-\eta^{2}\frac{2n}{\ell+1}+\eta^{4}\frac{2n\left(n-1\right)\left(\ell+1\right)+n^{2}\left(\ell+2\right)}{\left(\ell+1\right)^{2}\left(\ell+2\right)}+O\left(\eta^{6}\right)\right]
\end{align}
Note that, at leading order in $n^{-1}$, we just get $f=0$. Therefore,
we need to expand it to first order: 
\begin{align}
\frac{n!}{\left(n-\ell\right)!} & =n(n-1)\cdots(n-\ell+1)=n^{\ell}+n^{\ell-1}\left(-1-2-\cdots-\left(\ell-1\right)\right)+O\left(n^{\ell-2}\right)\\
 & =n^{\ell}-n^{\ell-1}\frac{\ell\left(\ell-1\right)}{2}+O\left(n^{\ell-2}\right)\\
\eta^{2m} & =\frac{\tilde{\eta}^{2m}}{\left(n+1/2\right)^{m}}=\frac{\tilde{\eta}^{2m}}{n^{m}}\left(1+\frac{1}{2n}\right)^{-m}=\frac{\tilde{\eta}^{2m}}{n^{m}}\left(1-\frac{m}{2n}+O\left(n^{-2}\right)\right)\\
\left(\ell!\right)^{2}f(n,\ell,\eta) & \approx\left(n^{\ell}-n^{\ell-1}\frac{\ell\left(\ell-1\right)}{2}\right)\Bigg[1-\frac{2\tilde{\eta}^{2}\left(1-\frac{1}{2n}\right)\left(1-\frac{\ell}{n}\right)}{\left(\ell+1\right)}\;+\\
&\;\;\;\;\;\;\;\;\;\;\;\;\;\;\;\;\;\;\;\;\;\;\;\;\;\;\;\;\frac{\tilde{\eta}^{4}\left(1-\frac{1}{n}\right)\left[2\left(\ell+1\right)\left(1-\frac{2\ell+1}{n}\right)+\left(1-\frac{2\ell}{n}\right)\left(\ell+2\right)\right]}{\left(\ell+1\right)^{2}\left(\ell+2\right)}+O\left(\eta^{6}\right)\Bigg]\;-\\
 & \;\;\;\;\left(n^{\ell}+n^{\ell-1}\frac{\ell\left(\ell+1\right)}{2}\right)\left[1-\frac{2\tilde{\eta}^{2}\left(1-\frac{1}{2n}\right)}{\left(\ell+1\right)}+\frac{\tilde{\eta}^{4}\left(1-\frac{1}{n}\right)\left[2\left(1-\frac{1}{n}\right)\left(\ell+1\right)+\left(\ell+2\right)\right]}{\left(\ell+1\right)^{2}\left(\ell+2\right)}+O\left(\eta^{6}\right)\right]\\
 & =n^{\ell-1}F\left(\ell,\tilde{\eta}\right)+O\left(n^{\ell-2}\right)
\end{align}
where we see explicitly that the term proportional to $n^{\ell}$
cancels out. Finally, we find the leading correction to the long-range interaction is
\begin{align}
g_\mathrm{eff}^{s} & \approx s\frac{J^{2}}{\omega}\left(\sum_{\ell_s}\left[\frac{n^{\ell-1}F\left(\ell,\tilde{\eta}\right)}{\ell\left(\ell!\right)^{2}}+O\left(n^{\ell-2}\right)\right]\eta^{2\ell}\right)\\
 & = s \frac{J^{2}}{n\omega}\left(\sum_{\ell_s}\frac{F\left(\ell,\tilde{\eta}\right)\tilde{\eta}^{2\ell}}{\ell\left(\ell!\right)^{2}}\right)+O\left(n^{-2}\right)
\end{align}
where $s=\pm 1$ tells us whether to sum over even ($s=1$) or odd ($s=-1$) terms. As expected, the leading correction at fixed $\tilde{\eta}$ goes
as $1/\left(n\omega\right)$ for large $n$. Notice that this vanishes
very quickly for $\ell\gg1$, so there is no need to consider any
limit besides $\ell\ll n$.

\section{Role of gauge freedom in many-body localization}

Most papers on cavity QED implicitly work in the dipole gauge yet
neglect the $P^{2}$ term. A generic such Hamiltonian can be written
\[
H_{\tilde{D}}^{gen}=\sum_{j}\left[\frac{p_{j}^{2}}{2m}+V({\bf r_{j}})-\omega q{\bf r}_{j}\cdot{\bf f}({\bf r}_{j})\left(a+a^{\dagger}\right)\right]+\omega a^{\dagger}a
\]
where we will use the convention that a tilde over the gauge label
indicates this missing term. Truncating this Hamiltonian to get a
Fermi-Hubbard model similar to that in the main text, one finds 
\begin{equation}
H_{\tilde{D}}=\sum_{j}\left[-J\left(c_{j}^{\dagger}c_{j+1}+h.c.\right)+Un_{j}n_{j+1}+V_{j}n_{j}-\omega\eta jn_{j}\left(a+a^{\dagger}\right)\right]+\omega a^{\dagger}a\label{eq:H_dipole_wrong}
\end{equation}
which seems natural. However, we will soon see that this incorrect
quantization leads to significant differences, including qualitative
changes in the ability to localize the system.

We can move to the Coulomb gauge by the same gauge transformation
as before, $U_{PZW}=\exp\left[-i\sum_{j}qdjn_{j}f_{0}\left(ia^{\dagger}-ia\right)\right]$.
But now we find 
\begin{equation}
H_{\tilde{C}}=\sum_{j}\left[-J\left(e^{-i\eta\left(ia^{\dagger}-ia\right)}c_{j}^{\dagger}c_{j+1}+h.c.\right)+Un_{j}n_{j+1}+V_{j}n_{j}\right]+\omega a^{\dagger}a-\omega\left(\sum_{j}\eta jn_{j}\right)^{2}.\label{eq:H_coulomb_wrong}
\end{equation}
In other words, the infinite range $P^{2}$ interaction appears in
the Coulomb gauge, albeit with opposite sign as in the correct quantization.
This is problematic for a number of reasons. First, it kills translation
invariance in this gauge, where it is expected on physical grounds.
Second, it remains untouched in the high-frequency expansion, which
must be done in the Coulomb gauge for strong drive, as we saw in the
main text. Therefore, even at zeroth order, one finds long-range interactions
which compete with hopping and local interactions: 
\[
H_\mathrm{eff}^{0}=\sum_{j}\left[-J_\mathrm{eff}\left(c_{j}^{\dagger}c_{j+1}+h.c.\right)+Un_{j}n_{j+1}+V_{j}n_{j}\right]-\omega\left(\sum_{j}\eta jn_{j}\right)^{2}
\]
Indeed, the existence of this last term calls the high-frequency expansion
into question, as it is large (proportional to $\omega$) yet has
not been made diagonal by the canonical transformation.

The effects of this term also show up numerically. As seen in Figure
\ref{fig:mbl_wrong_hamiltonian}, the system shows little sign of
localization in the presence of this additional term, even in the
case where one has nearly perfect coherent destruction of tunneling
($E/\omega\approx n_{t}$). This seems to be due to the additional
long-range interactions, which make localization fragile.

\begin{figure}
\centering \includegraphics[trim=0.8cm 0.6cm 1.0cm 4.6cm,clip,width=0.98\columnwidth]{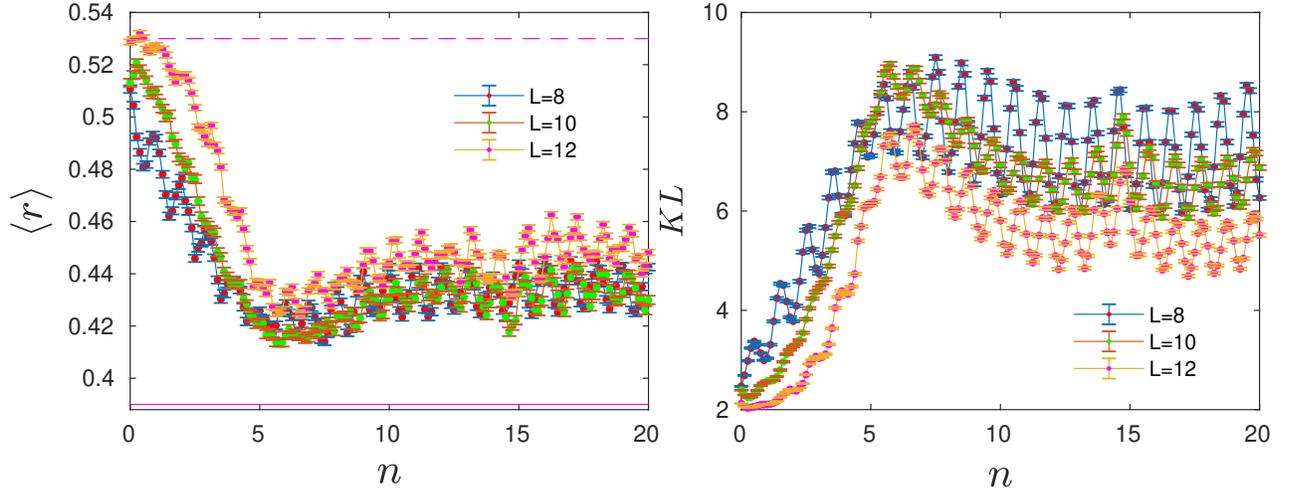}
\caption{Level statistics (left) and Kullback-Leibler divergence (KL) (right)
for $n_{t}=6$ when using the incorrect quantization, shown in Eqs.~\ref{eq:H_dipole_wrong}
and \ref{eq:H_coulomb_wrong}. Parameters are similar to those in
Figure \ref{fig:mbl_full_ed}, except that $n_{t}=5.523$ ($\eta=0.49$).}
\label{fig:mbl_wrong_hamiltonian} 
\end{figure}

}
\end{document}